\documentclass{aa}
\usepackage[varg]{txfonts}
\usepackage{graphicx}
\usepackage{natbib}
\bibpunct{(}{)}{;}{a}{}{,} 

\begin{document}

\titlerunning{Solar activity during the Maunder minimum}
\authorrunning{Vaquero et al.}

\title{Level and length of cyclic solar activity during the Maunder minimum as deduced from the active day statistics}

\author{J.M. Vaquero\inst{1}
\and G. A. Kovaltsov\inst{2}
\and I.G. Usoskin\inst{3}
\and V.M.S. Carrasco\inst{4}
\and M.C. Gallego\inst{4}
}

\institute{
Departamento de F\'isica, Universidad de Extremadura, M\'erida, Spain
\and Ioffe Physical-Technical Institute, 194021 St. Petersburg, Russia
\and Sodankyl\"a Geophysical Observatory and ReSoLVE Center of Excellence, University of Oulu, Finland
\and Departamento de F\'isica, Universidad de Extremadura, Badajoz, Spain
}

\date{}

\abstract {}
{The Maunder minimum (MM) of greatly reduced solar activity took place in 1645--1715, but the exact level of sunspot activity
 is uncertain as based, to a large extent, on historical generic statements of the absence of spots on the Sun.
Here we aim, using a conservative approach, to assess the level and length of solar cycle during the Maunder minimum,
 on the basis of direct historical records by astronomers of that time. }
{A database of the active and inactive days (days with and without recorded sunspots on the
 solar disc respectively) is constructed for three models of different levels of conservatism (loose ML, optimum MO and
 strict MS models) regarding generic no-spot records.
We have used the active day fraction to estimate the group sunspot number during the MM.
}
{A clear cyclic variability is found throughout the MM with peaks at around 1655--1657, 1675, 1684 and 1705, and possibly
 1666, with the active day fraction not exceeding 0.2, 0.3 or 0.4 during the core MM, for the three models.
Estimated sunspot numbers are found very low in accordance with a grand minimum of solar activity.
 }
{We have found, for the core MM (1650-1700), that:
(1) A large fraction of no-spot records, corresponding to the solar meridian observations, may be unreliable in the conventional database.
(2) The active day fraction remained low (below 0.3--0.4) throughout the MM, indicating the low level of sunspot activity.
(3) The solar cycle appears clearly during the core MM.
(4) The length of the solar cycle during the core MM appears $9\pm 1$ years, but there is an uncertainty in that.
(5) The magnitude of the sunspot cycle during MM is assessed to be below 5--10 in sunspot numbers;
 A hypothesis of the high solar cycles during the MM is not confirmed.
}

\maketitle

\section{Introduction}

There was a period, in the second part of the 17th century, of greatly reduced solar activity that was named the Maunder Minimum (MM) by \citet{eddy76}.
The MM was characterized by almost complete absence of reported sunspots on the solar surface although some indications of
 cyclic activity can be noticed particularly in the geomagnetic and heliospheric indices \citep{beer98,usoskin_JGR_MM_01,soon03}.
The reconstruction of solar activity based on the historical records of telescopic observations of sunspots since 1610 \citep[][ -- called HS98 henceforth]{hoyt98,hoyt98a}
 marked a milestone in the study of solar activity in the recent past and, especially, for the MM period.
The Group Sunspot Number (GSN) built by HS98 became the only high-resolution (daily) index to study solar activity during the MM.

The aim of this work is to elucidate whether the absence of the sunspot cyclic activity during the MM was real or an artefact caused by a problem of compilation
 of the database of sunspot records.
Several studies pointed to possible inconsistences in the database used by HS98 especially around the MM \citep[e.g.,][]{vaquero09,vaquero11,vaquero_SP_14}.
As an extreme, \citet{zolotova15} claimed there was no grand Maunder minimum and that sunspot cycles during MM were as high as $\approx 100$,
 which is higher than the current cycle \# 24.
We note that the MM is well covered by sunspot data and more than 90\% of days have formal observation records in the HS98 database.
However, it contains a large number of generic statements of the absence of sunspots during a long period of time.
Such records are not strict observational data but they were interpreted by HS98 as no-spot data.
Many of these records corresponded to solar meridian observations \citep{vaquero_rev_07,clette14} and should be used with caution for
 the reconstruction of solar activity, as shown by \citet{vaquero_ASR_14} who analyzed sunspot records taken during systematic solar meridian observations performed
 at the Royal Observatory of the Spanish Navy from 1833 to 1840.
Moreover, as \citet{carrasco15} suggested basing on an analysis of sunspot records by Hevelius in the 17th century,
 the GSN index may be underestimated during the MM due to a large number of "zero" sunspot records taken from solar meridian observations.
In general, astrometric observations of the Sun are not always reliable for sunspot counting because of the different aim of such observations.
For example, there is no information on sunspots in the extensive table of astrometric records of the Sun made with the meridian line in the San Petronio
 Basilica from 1645 to 1735 as published by \citet{manfredi1736}.
Nevertheless, \citet{hoyt98} adopted solar observations recorded in this source as no-spot reports, which is not correct.
It has been discussed that, while the definition of sunspot numbers and even sunspot groups is not very reliable in the earlier part of the GSN series \citep{clette14,zolotova15},
 solar activity during the MM can be reliably represented by the fraction of active days \citep{kovaltsov04, vaquero12,vaquero_ASR_14, usoskin_LR_13}.

Despite the overall level of activity, the parameters of the solar cyclic variability during MM are also important to know.
Although the solar cycle was perceptible in the butterfly diagram \citep{ribes93,vaquero_ASR_15} based on the observations of sunspot latitudes
 during the last decades of the 17th century, the 11-year solar cycle is only marginally detectable in the sunspot numbers \citep{waldmeier61,mendoza97}
 with a weak 22-year cycle dominated \citep{usoskin_JGR_MM_01}.
On the other hand, some works based on data of high-resolution cosmogenic $^{14}$C measured in tree trunks suggest that the solar cycle might have been stretched
 during Grand solar minima \citep{stuiver98,miyahara04, miyahara06, miyahara10, nagaya12,miyake13}.
These studies have suggested that the length of the solar cycle was increased to about 14 years during the MM, to about 13 years in the beginning
 of the Sp\"orer Minimum, up to 16 years during the 4th Century BC Minimum, and to 12--13 years during the late 7th century minima.

In this work we aim to study variability of solar activity during the MM using the statistics of the active days basing on only the most reliable solar observations
 from the database compiled by \citet{hoyt98} and to establish an uppermost upper ({\it maximum maximorum}) limit on that.

\section{Sunspot activity database}

Since quantitative interpretation of many records is uncertain for that period, we consider only qualitative indicators of the sunspot activity for each day
 for the period 1637--1715 AD.
Leaving aside the exact number of reported sunspot group in the HS98 catalog, we only consider three possible states for each day:
\begin{itemize}
\item
 no-information or missing days;
\item
 inactive days when we believe there were reliable observation of the absence of sunspots;
\item
 active days when at least one sunspot group was explicitly reported by at least one observer;
\end{itemize}
We build our database of the active and inactive days for three models of different levels of conservatism regarding generic no-spot records.
For the period 1637--1642, we used exactly the records listed by \citet{vaquero11}.
For the period 1643--1715, we used the records from the HS98 database (http://www.ngdc.noaa.gov/stp/sunspot\_numbers/group\_ sunspot\_numbers/alldata.dat) for each observer separately.
In addition, for the year 1672 several active days were added according to observations by N. Bion not included into the HS98 database \citep{casas06}.
While the original HS98 database contains 26508 daily records for the analyzed period 1637--1715, our models include much less
 records because of rejecting, with different levels of conservatism, generic statement mostly related to no-spot observations.
All these models provide an overestimated upper bound of sunspot activity due to a possible selection bias towards active days.

\subsection{Loose model (ML)}

This model is similar to that by \citet{kovaltsov04} and ignores all the generic statements (longer than a month) in the HS98 database, and considers only explicit
 statements with exact mentioning dates of observations.
This affects such generic statements as, e.g., by J. Hevelius for no spots during 1645--1651, by J. Picard for 1653--1665, by H. Siverius for 1675-1689, etc.
This model is least conservative and is called "loose".
It includes 13512 observational days which is nearly half of the HS98 database.

\subsection{Optimum model (MO)}

The MO model provides a reasonable balance between strictness and data acceptance and is considered as the optimum conservative.
For each year, we considered observations of only those observers who reported at least one sunspot group at any day of the year, which would
 prove that the observer was "active".
In this way, generic long-extending reports of "no-spot" were neglected but no-spot records of active observers were considered.
The MO models is biased towards "active" years and produces no result for the years without sunspot observations.
For example, if a year is full of definite "no-spot" records but does not contain a single sunspot observation reported,
 such year is marked as "no-information" in this model.
Alternatively, if an observer was "active" during a year, his generic "no-spot" records for this year were considered by the model,
 so that the total number of days $N_T$ in the MO model may exceed that for the ML model for some years (see Fig.~\ref{Fig:days}).
This model includes 8089 observational days for the period analyzed, which is roughly $^1$/$_3$ of the full HS98 database.

\subsection{Strict model (MS)}

In this "strict" model we excluded all the generic statements as in the ML models, but additionally we treated other no-spot records in a very
 conservative way, so that we consider as inactive only days, when at least two observers independently reported that the Sun was spotless and there were
 no other records of sunspots.
If at least one observer reported sunspots, the day was considered as active.
All other days were treated as no-information days.
This is the most conservative approach, especially in the earlier part of the Maunder minimum, when the number of documented observers
 was low and they rarely overlap.
This model includes 5159 daily records or $^1$/$_5$ of the full HS98 database.

\begin{figure}
\begin{center}
\resizebox{\columnwidth}{!}{\includegraphics{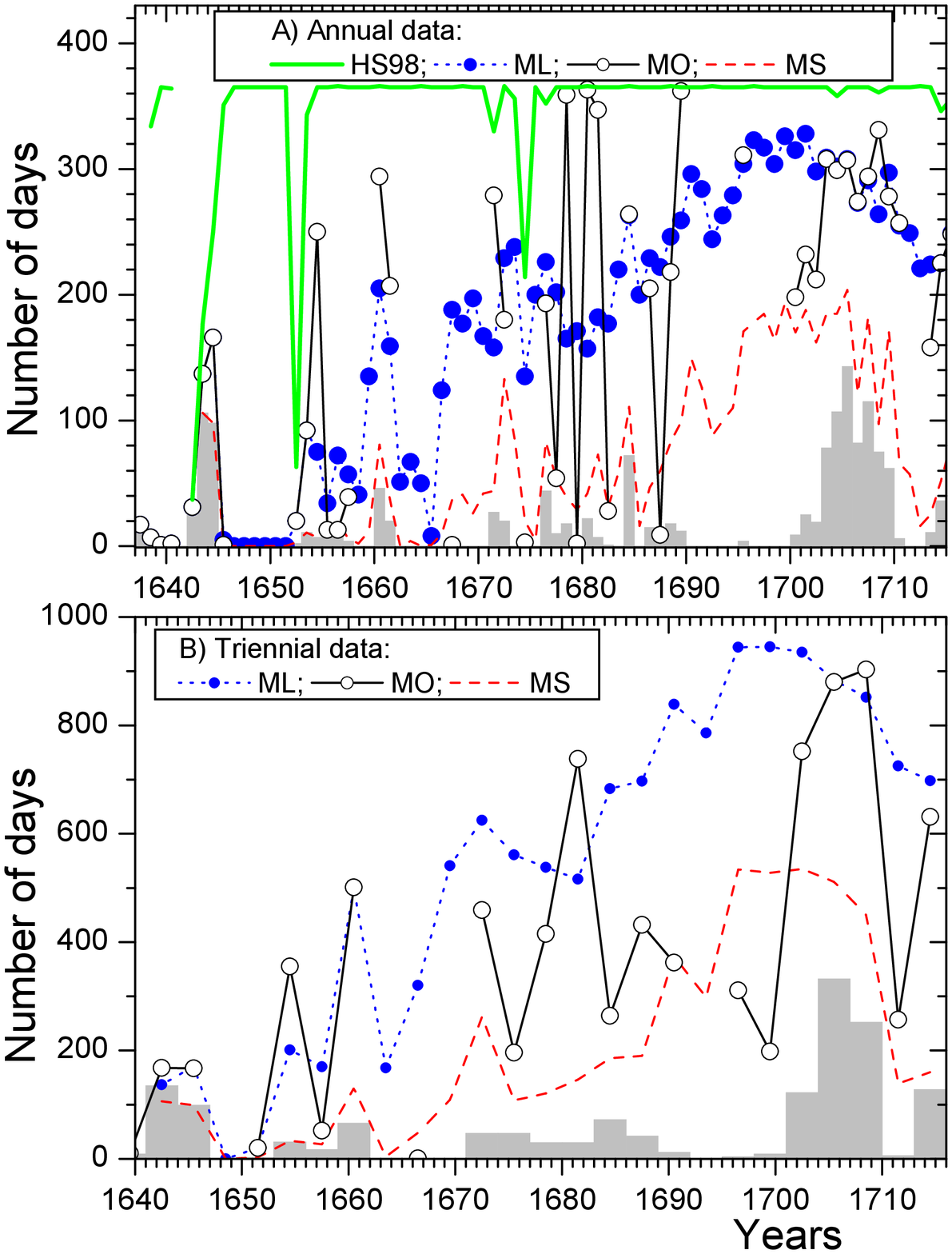}}
\end{center}
\caption{ The number of active days $N_A$ (grey bars) and total observational days $N_T$ (curves as defined in the legend for
 the three models and the formal HS98 database) per interval used.
 Panels A and B depict annual and triennial data, respectively.
\label{Fig:days}}      
\end{figure}

For each model we define the number of active $N_A$ and the total number $N_T$ of the accepted observational days in a year.
Since the annual data are quite noisy (see below) we also consider triennial intervals.
In order to keep the strictness, the MO model was still operating with annual periods to identify "active" observers.
The results are shown in Fig.~\ref{Fig:days} for the three models as well as for the formal HS98 database.
One can see that, while the HS98 database covers the entire period pretty well, the three models provide a more conservative estimate
 of reliable observations, which is greatly reduced in the earlier Maunder minimum but is quite solid towards its end.

\begin{figure}
\begin{center}
\resizebox{\columnwidth}{!}{\includegraphics{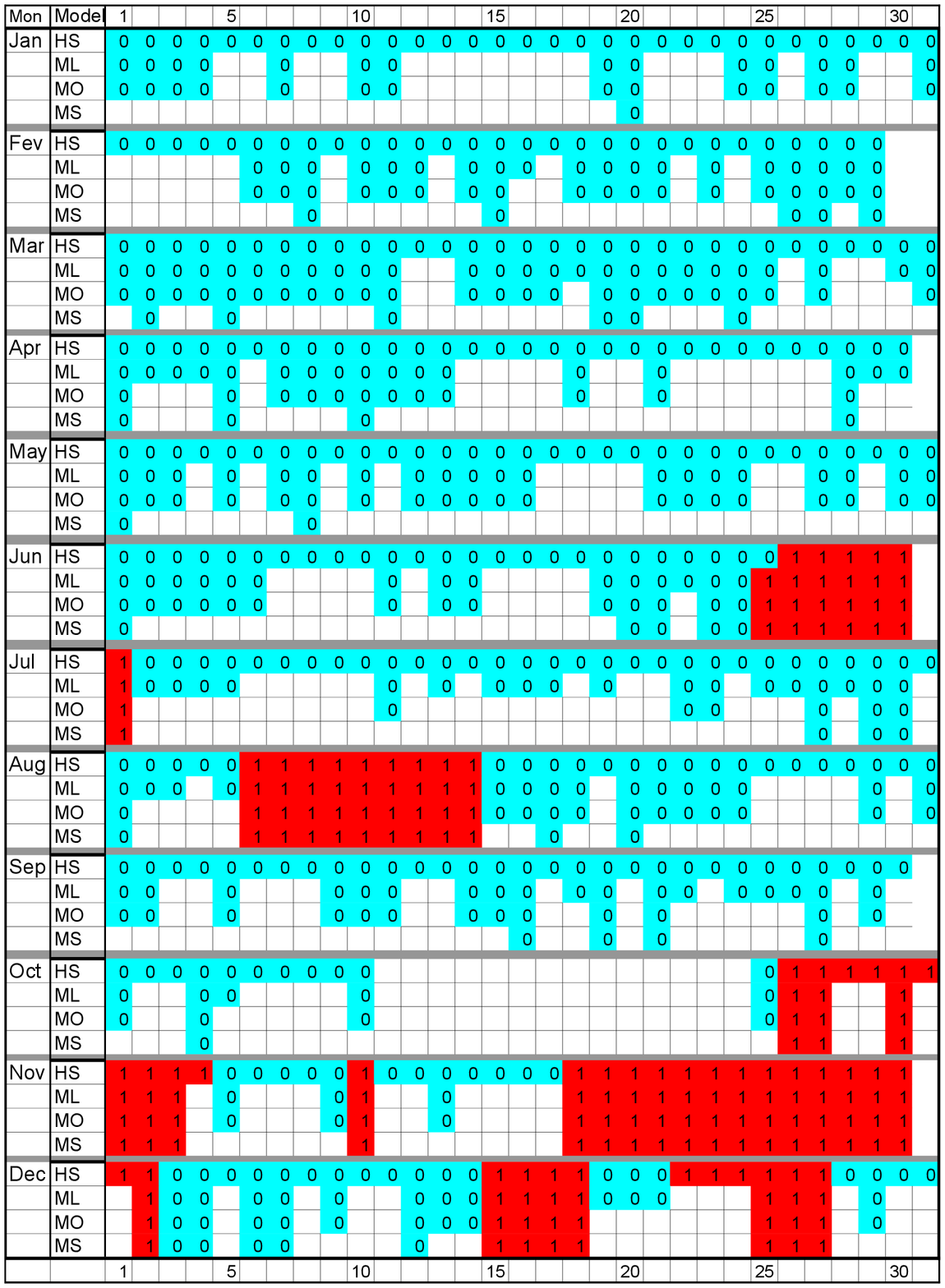}}
\end{center}
\caption{ A map of the days allocation for the year 1676 in the original HS98 database (denoted as HS) and
 the three models considered here.
 Each line represents one month (days of the month numbered on the top and bottom) for a model.
 The empty white, blue "0" and red "1" cells correspond to no-information, no-spot and active days.
\label{Fig:Y1676}}      
\end{figure}

An example of the coverage of the data in the three models and the formal HS98 database is shown in Fig.~\ref{Fig:Y1676} for the year 1676.
Although this year was almost fully covered by data in the formal HS98 database, except for a short gap in October, the three models considered
 here include much less inactive days while keeping the active days.
A small discrepancy in the number of active days is related to excluded interpolations (as in Dec 22--24) and confusing values (as in Jun 25 when a sunspot record
 by R. Hook was missed in the formal HS98 series) in the HS98 database.

We emphasize again that the procedure described above serves as an uppermost upper bound because of possible over-suppressing zero-sunspot records.

\section{Results}

\subsection{Active day fraction}

\begin{figure}
\begin{center}
\resizebox{\columnwidth}{!}{\includegraphics{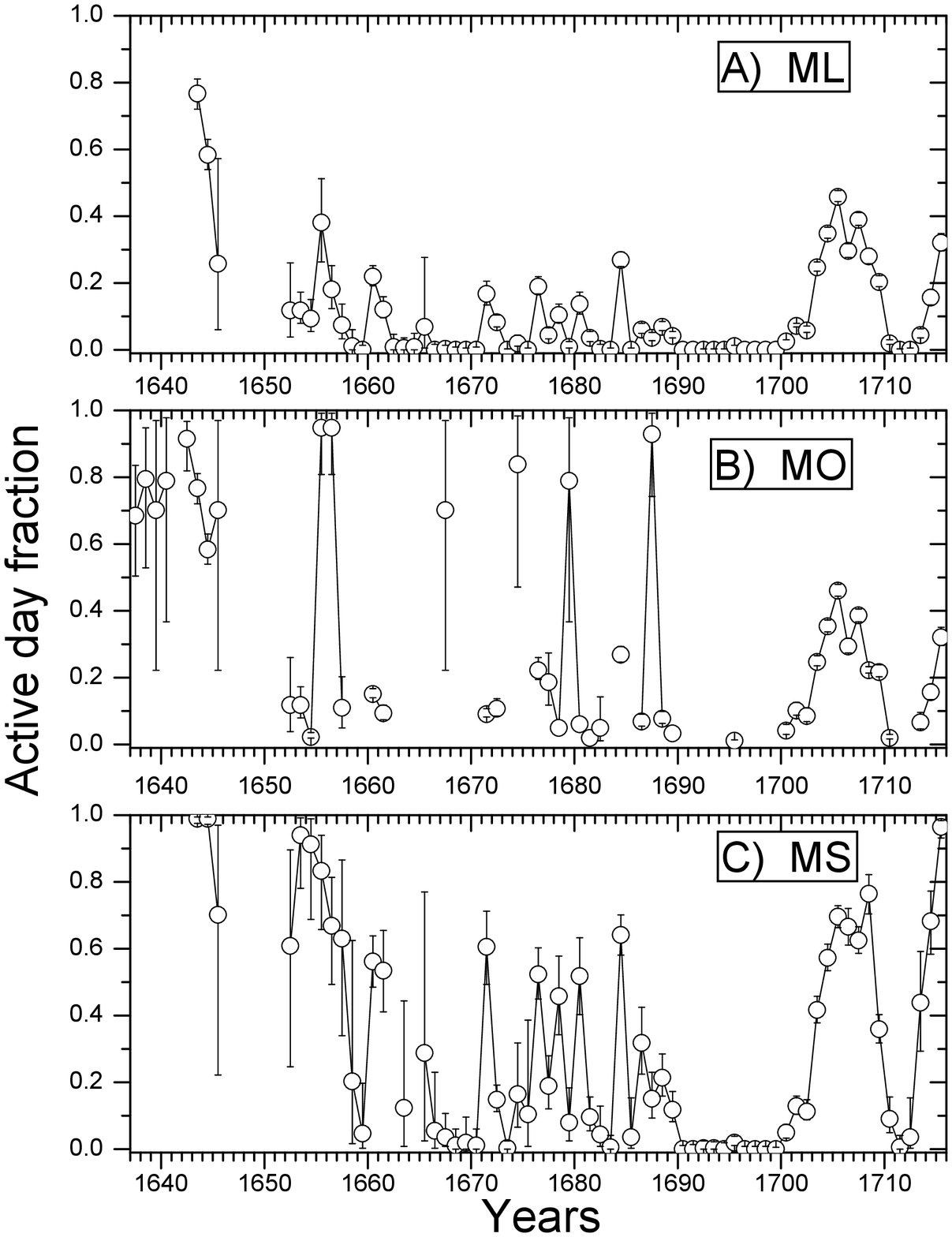}}
\end{center}
\caption{ Annual active day fraction for the three models.
 Error bars represent the 90\% two-sided uncertainties.
\label{Fig:AF1}}      
\end{figure}

\begin{figure}
\begin{center}
\resizebox{\columnwidth}{!}{\includegraphics{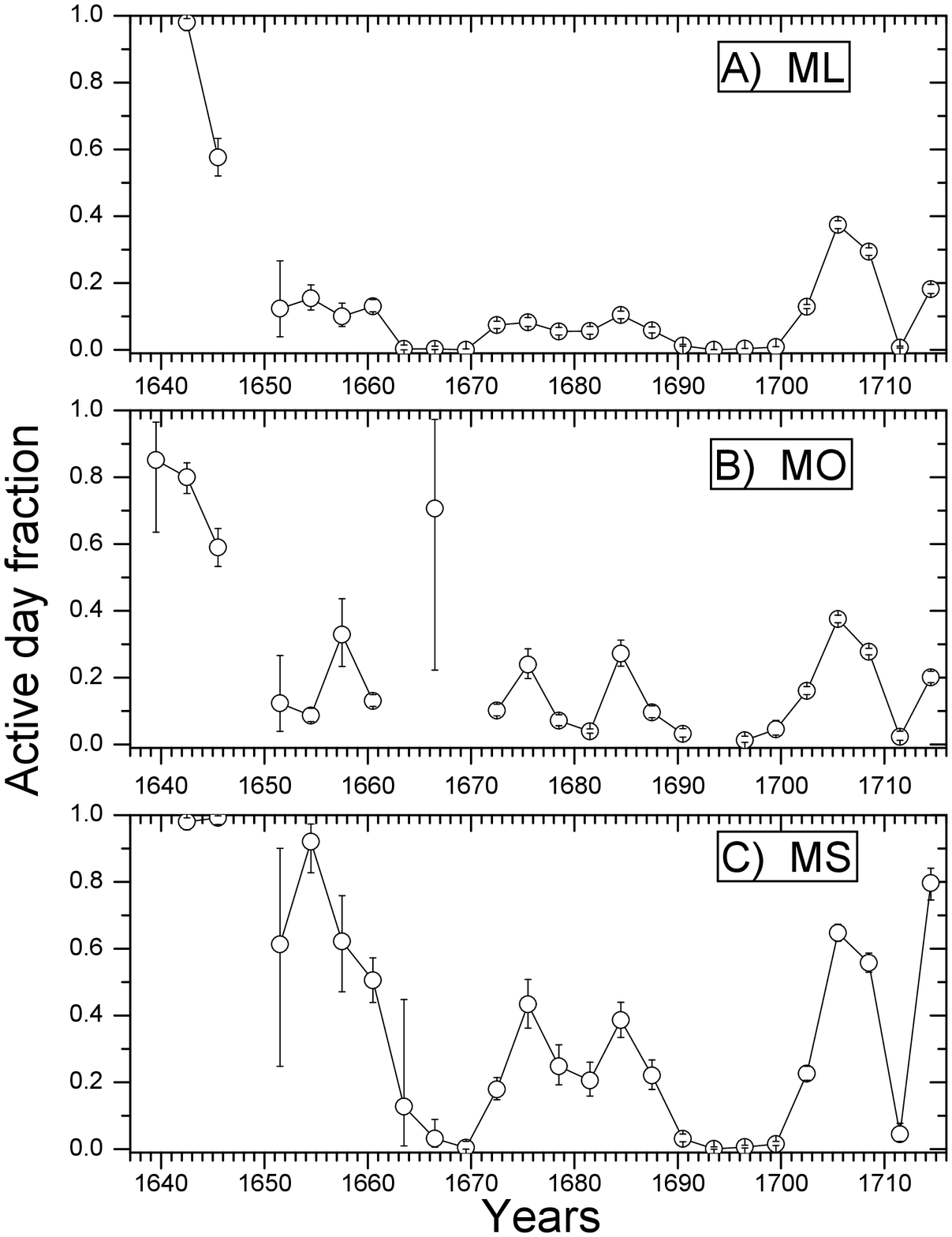}}
\end{center}
\caption{ Triennial active day fraction for the three models.
 Error bars represent the 90\% two-sided uncertainties.
 Digital data is available in Table~\ref{Tab:3y}.
\label{Fig:AF3}}      
\end{figure}

From the collected database of sunspot records, we have estimated the fraction of active days $F_A$ in each model, as follows \citep[cf.][]{kovaltsov04}.
For each interval, either annual or triennial, we have a sample of $n$ daily observations with $r$ active days reported.
Assuming these observation were taken randomly and independently, one can assess the probability of the occurrence of
 exactly $s$ active days within $N$ days during the considered interval (a year or 3 years)
 using the hypergeometric probability distribution:
\begin{equation}
p(s)={s!\,(N-s)!\over (s-r)!\,(N-s-n+r)!}\,\cdot\, {n!\,(N-n)!\over (n-r)!\, N!\, r!}
\label{Eq:1}
\end{equation} 						
As the optimum value of $s*$ we consider the median value, viz. the value of $s$ which yields $P(s*)\equiv \sum_{r}^{s*}{p(s)}=0.5$.
The results for annual and triennial time intervals are shown, along with error bars of a 90\% (two-sided) confidence interval,
 in Figures~\ref{Fig:AF1} and \ref{Fig:AF3}, respectively.
We note that triennial data were calculated from the original daily values using equation (\ref{Eq:1}) and not as an
 average of the annual data.

\subsection{Length of solar cycles}

Although the annual data are quite noisy, the triennial ones (see Table~\ref{Tab:3y}) clearly show a decadal periodicity during the MM.
For example, Figure~\ref{Fig:AF3} suggests  maxima of solar cycles around 1639, 1655--1657, 1675, 1684 and 1705
 in all the models.
There is also an indication of a cycle maximum around 1666 in the MO models, but the statistics is low with a single
 observation for the 3-year interval.
Periods around 1648 and 1693 are poorly known with data gaps in the MO model.

There are four solar activity maxima in the core MM, between maxima ca. 1657 and 1684.
This leads to an estimate of the average solar cycle length (max-to-max) during the core MM as $9\pm 1$ years.
However, our view of the cyclic evolution of sunspot activity during MM is uncertain because of the
 unclear situation around 1648, 1666, and 1693.
If we assume two hypothetical missing solar maxima during these periods, as e.g., \citet{waldmeier61} proposed
 a cycle maximum in 1649, while \citet{usoskin_JGR_MM_01} suggested a maximum ca. 1695,  we can estimate an average solar
 cycle length around the MM (from 1636 to 1711) to be $9.5\pm 0.5$ years.
If however, we assume that there were no additional solar cycle maxima around 1648 and 1693, the average cycle
 length (max-to-max) would be $13.2\pm 0.6$ years.
However in this case, the length of individual cycles varies greatly, between 9 and 18 years.
The estimated cycle length is similar to but somewhat shorter than the results proposed by \citet{mendoza97} and \citet{usoskin_JGR_MM_01}
 who suggested the cycle length of 10.5-11 years during the MM using sunspot observations.
Meanwhile, clustering of activity in $\approx 20-$year intervals (1650--1670, 1670--1690, and 1690--1710) is also
 visible, in agreement with earlier results of the dominant 22-year periodicity during the MM \citep{usoskin_JGR_MM_01}.
Note, however, that this clustering of activity could be also produced because of the scarcity of reliable data around
 1648, 1669, and 1693.

On the other hand, estimates of the cycle length based on cosmogenic $^{14}$C data suggest much longer cycles
 during Grand minima (13-16 years).
We note however that $^{14}$C data cannot resolve individual cycles, because of the global carbon cycle attenuating
 high-frequency variability \citep{roth13}, but rather yields the mean periodicity over the interval analyzed \citep[e.g.][]{miyahara04}.
This seeming contradiction between the results obtained here \citep[cf.][]{mendoza97,usoskin_JGR_MM_01} and from $^{14}$C data can be
 potentially reconciled in a view of the possible inversion of the cycle phase in the cosmic ray modulation during the periods of very weak
 activity like the MM \citep{owens12}.
Thus, one or two cycles can be lost in the $^{14}$C data, due to forward and then reverse phase shifts in the beginning and
 end of the Maunder minimum, leading to a seemingly extended cycles in $^{14}$C data.

\subsection{Sunspot numbers}

\begin{figure}
\begin{center}
\resizebox{\columnwidth}{!}{\includegraphics{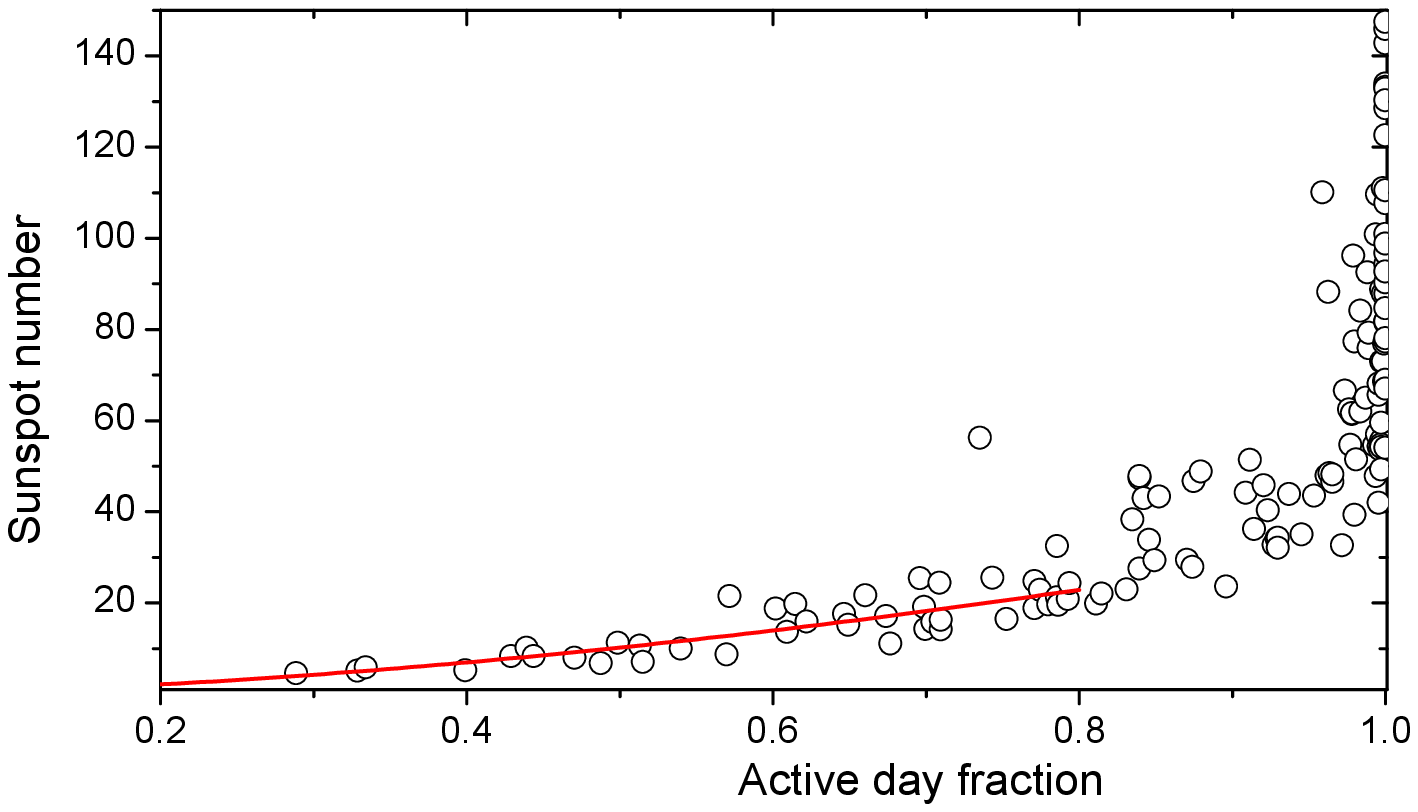}}
\end{center}
\caption{ Relation between triennial sunspot numbers and active day fraction for the period 1850--1995 using the
 Group Sunspot Number \citep{hoyt98}.
 The red curve is the best fit relation $R = 33.6\cdot F_{A}^{1.72}$.
\label{Fig:AFSN}}      
\end{figure}

On average, the fraction of active days observed during MM was low, below 0.4 in the triennial data (Fig.~\ref{Fig:AF3}) for ML and MO models,
 except for the year 1666 (MO model) which is however based on a single observation, and reaching up to $>0.7$ in the most conservative MS model.
We note that, for the normal cycles, the active day fraction is about 100\% except for the years around solar minimum \citep{kovaltsov04, vaquero12, vaquero_ASR_14}.
The value of $F_A$ was never below 0.15 for annual and 0.29 for triennial (see Fig.~\ref{Fig:AFSN}) during the period 1850--1995.
Accordingly, such low values $F_A$ even for the peaks during the MM correspond to (or are lower than) the minimum state of modern solar cycles.
Therefore, although a cyclic activity during the MM is clear, at least during the core period, the sunspot cycles were weak, with the maxima being comparable
 to the modern cycle minima.
We note that high solar cycles of the magnitude 40-100 in sunspot number as proposed by \citet{zolotova15} would
 unavoidably imply $\approx 100$\% active day fraction \citep{vaquero_ASR_14} during most of the years, which contradicts with the data
  (cf. Fig.~\ref{Fig:Y1676}).

In order to assess the sunspot number $R$ from the active day fraction $F_A$, we apply a method adopted from \citep{kovaltsov04, vaquero12, vaquero_ASR_14}.
For the annual data the relation was \citep{kovaltsov04}:
$R=19\cdot F_{A}^{1.25}$ for $F_{A}\leq 0.5$ and $R=2.1\cdot\exp{(2.69\cdot F_A)}$ for $0.5<F_A\leq 0.8$.
The relation between triennial values $R$ and $F_{A}$ is shown in Figure~\ref{Fig:AFSN} for the period 1850--1995.
One can see that the relation is quite good for $F_A<0.8$ (with the only outlier related
 to the period 1954--1956 which corresponded to the growth phase of the highest solar cycle \#19) and can be
 well approximated by a dependence $R=33.6\cdot F_{A}^{1.72}$.
The relations loosens for $F_{A}>0.8$ and is lost completely with the
 active day fraction approaching unity.
Thus, the active day fraction is a good index of sunspot activity until it reaches 0.8.

\begin{figure}
\begin{center}
\resizebox{\columnwidth}{!}{\includegraphics{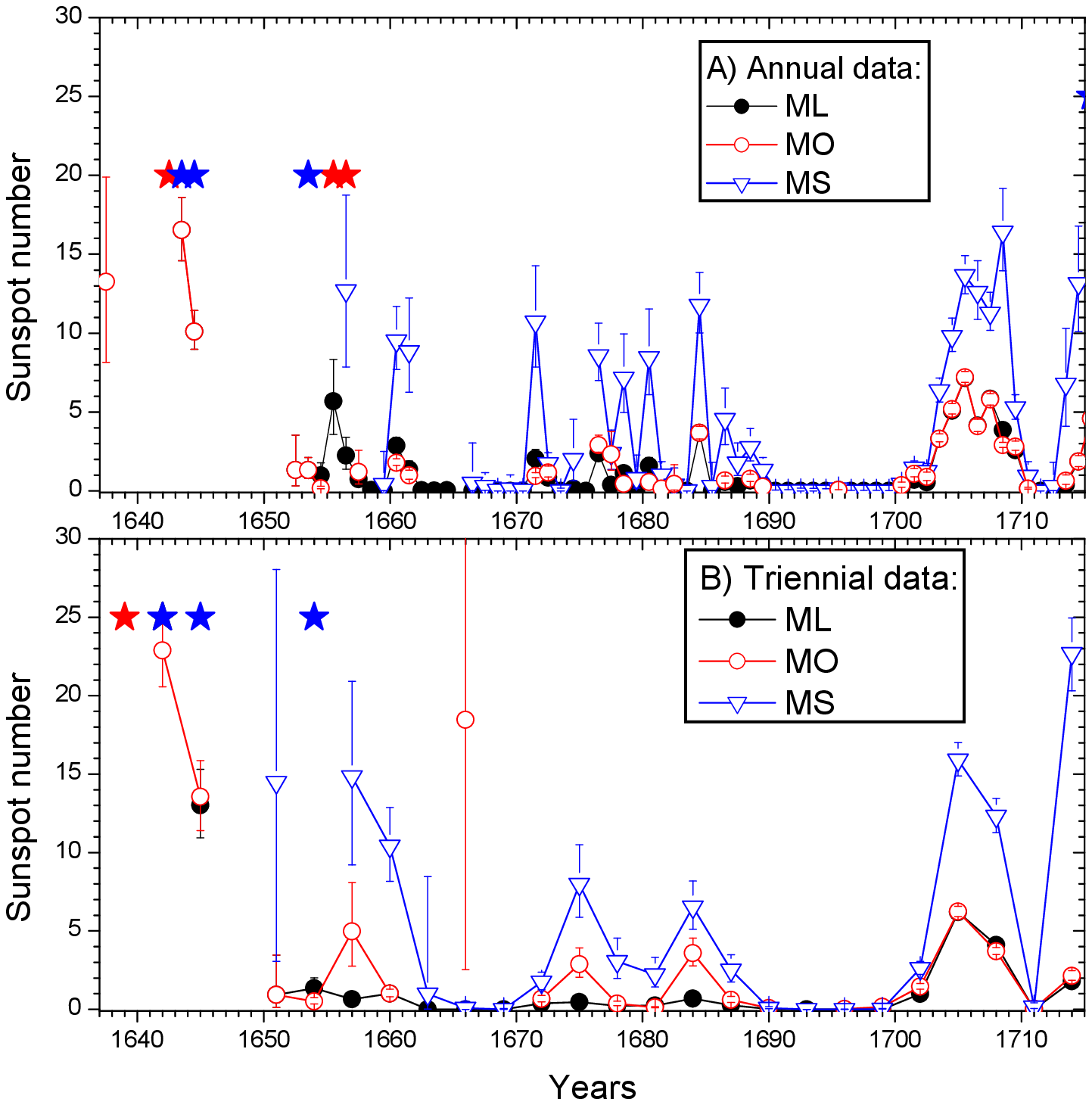}}
\end{center}
\caption{ Annual (panel A) and triennial (panel B) sunspot numbers reconstructed in the three models as denoted in the legends.
Years with low statistics ($N_T<10$) are not shown.
Stars indicate that the sunspot number cannot be assessed from the active day fraction (see text) and is greater than 18/23 for the
 annual/triennial data.
\label{Fig:SN3}}      
\end{figure}

 Using these dependencies we have evaluated the sunspot numbers during the period analyzed, as shown in Figure~\ref{Fig:SN3}.
One can see that the sunspot numbers appear below 2 during the deep MM (1645--1700) and 7 ca. 1705 in the least conservative model ML.
The optimum MO model yields the sunspot number not exceeding 5 for the deep MM and 7 ca. 1705 (except for the very uncertain period ca. 1666
 with the lack of observations).
The most conservative MS model yields sunspot cycles below 10 during the core MM and a possible relatively high cycle
 in the 1650s, which is based on the lack of overlapping records from different observers, and about 15 ca. 1705.
Anyway, even these very strict model suggests that the cycles were lower than 15--20 in sunspot numbers, which is much lower
 than the present cycle \#24 and an order of magnitude lower than the very high cycles proposed by \citet{zolotova15}.
Considering the severe reduction of the statistics and a possible strong bias towards active days in the MS model, we
 believe it is not indicative for the true solar activity evolution during the MM and may represent only the
 uppermost upper ({\it maximum maximorum}) bound.

\section{Conclusions}

Using three models of different level of conservatism to treat generic "no-sunspot" statements, we have created a database of reliable
 sunspot observation around the Maunder minimum (1637-1715) and revised the sunspot cyclic activity over that period.
 We show that:
\begin{enumerate}
\item A large number of no-spot records, corresponding to the solar meridian observations, may be unreliable in the HS98 database.

\item
The active day fraction remained low (below 0.3--0.4) throughout the MM, indicating the low level of sunspot activity.

\item The solar cycle appears clearly during the core MM with maxima at 1657, 1675, 1684, 1705 and possibly 1666.

\item The length of the solar cycle during the MM appears shorter ($9\pm 1$ years) in comparison with the standard 11-year solar cycle, but there is
 an uncertainty in that.
 A $\approx 20-$year clustering of activity is also observed.

\item The magnitude of the sunspot cycle during MM is assessed to be below 5 (10 in the most conservative model) in sunspot numbers.
The exact level is hardly possible to determine but it is below 10.

\item High solar cycles during the Maunder minimum, as proposed by \citet{zolotova15}, contradict with the data.
\end{enumerate}

We note that this is an uppermost upper ({\it maximum maximorum)} bound for solar activity during MM because of a possible selection bias (particularly
 important in the MS model), and the true level of activity may be smaller than that.

In any case, only a thorough review of each record and each solar observation during the MM can make it possible to reveal
 the best picture of solar activity during this period.
Therefore, we encourage researchers (especially Latin scholars) to query and analyze the old texts to understand how
 the observations were made and the true level of solar activity they indicate.

\begin{table*}
\caption{Triennial statistics of sunspot day occurrence for the three models considered here (see text for definition).
Columns are:
\#1 - central year of the triennial interval;
\#2 - number of active days  N$_{A}$ within the interval;
\#3, 7 and 11 - number of total observational days N$_{T}$ considered in the three models, respectively;
\#4, 8 and 12 - lower 90\% bound of the active day fraction, for the tree models, respectively;
\#5, 9 and 13 - median active day fraction, for the tree models, respectively;
\#6, 10 and 14 - upper 90\% bound of the active day fraction, for the tree models, respectively. }
\begin{tabular}{l|l|lccc|lccc|lccc}
\hline
 &  & \multicolumn{4}{c|}{ML} & \multicolumn{4}{c|}{MO} & \multicolumn{4}{c}{MS}\\
Year & N$_{A}$ & N$_{T}$ & F$_{\rm low}$ & F$_{\rm med}$ & F$_{\rm up}$ & N$_{T}$ & F$_{\rm low}$ & F$_{\rm med}$ & F$_{\rm up}$ & N$_{T}$ & F$_{\rm low}$ & F$_{\rm med}$ & F$_{\rm up}$ \\
\hline
1639 & 9 & 10 & 0.636 & 0.851 & 0.965 & 10 & 0.636 & 0.851 & 0.965 & 9 & 0.742 & 0.932 & 0.994\\
1642 & 135 & 137 & 0.956 & 0.980 & 0.992 & 168 & 0.752 & 0.800 & 0.843 & 106 & 0.956 & 0.980 & 0.992\\
1645 & 99 & 171 & 0.521 & 0.576 & 0.633 & 167 & 0.533 & 0.590 & 0.647 & 99 & 0.972 & 0.992 & 0.998\\
1648 & 0 & 0 & N/A & N/A & N/A & 0 & N/A & N/A & N/A & 0 & N/A & N/A & N/A\\
1651 & 2 & 20 & 0.039 & 0.123 & 0.267 & 20 & 0.039 & 0.123 & 0.267 & 3 & 0.248 & 0.613 & 0.900\\
1654 & 31 & 201 & 0.120 & 0.154 & 0.195 & 355 & 0.068 & 0.087 & 0.109 & 33 & 0.827 & 0.921 & 0.974\\
1657 & 17 & 170 & 0.070 & 0.100 & 0.140 & 52 & 0.234 & 0.329 & 0.437 & 27 & 0.471 & 0.622 & 0.759\\
1660 & 66 & 499 & 0.114 & 0.131 & 0.151 & 501 & 0.114 & 0.131 & 0.150 & 130 & 0.439 & 0.506 & 0.573\\
1663 & 0 & 168 & 0 & 0.003 & 0.015 & 0 & N/A & N/A & N/A & 4 & 0.010 & 0.128 & 0.448\\
1666 & 1 & 320 & 0.001 & 0.003 & 0.011 & 1 & 0.223 & 0.706 & 0.974 & 48 & 0.006 & 0.032 & 0.089\\
1669 & 0 & 541 & 0 & 0 & 0.003 & 0 & N/A & N/A & N/A & 109 & 0 & 0.005 & 0.024\\
1672 & 47 & 625 & 0.064 & 0.074 & 0.086 & 459 & 0.086 & 0.101 & 0.121 & 262 & 0.148 & 0.179 & 0.215\\
1675 & 47 & 561 & 0.070 & 0.082 & 0.097 & 196 & 0.197 & 0.239 & 0.287 & 108 & 0.363 & 0.434 & 0.509\\
1678 & 30 & 538 & 0.045 & 0.055 & 0.068 & 415 & 0.057 & 0.071 & 0.089 & 121 & 0.193 & 0.248 & 0.312\\
1681 & 30 & 516 & 0.047 & 0.057 & 0.070 & 738 & 0.034 & 0.039 & 0.047 & 146 & 0.159 & 0.205 & 0.260\\
1684 & 72 & 683 & 0.093 & 0.104 & 0.116 & 264 & 0.235 & 0.272 & 0.312 & 186 & 0.334 & 0.386 & 0.440\\
1687 & 42 & 697 & 0.051 & 0.058 & 0.068 & 432 & 0.080 & 0.096 & 0.116 & 190 & 0.179 & 0.221 & 0.268\\
1690 & 12 & 839 & 0.011 & 0.013 & 0.016 & 362 & 0.022 & 0.032 & 0.047 & 374 & 0.022 & 0.031 & 0.046\\
1693 & 0 & 786 & 0 & 0 & 0.001 & 0 & N/A & N/A & N/A & 298 & 0 & 0.001 & 0.007\\
1696 & 4 & 944 & 0.004 & 0.004 & 0.005 & 311 & 0.006 & 0.013 & 0.025 & 534 & 0.004 & 0.006 & 0.012\\
1699 & 9 & 945 & 0.008 & 0.008 & 0.010 & 198 & 0.028 & 0.046 & 0.072 & 528 & 0.011 & 0.016 & 0.024\\
1702 & 122 & 935 & 0.123 & 0.129 & 0.136 & 752 & 0.150 & 0.161 & 0.174 & 535 & 0.206 & 0.226 & 0.248\\
1705 & 332 & 883 & 0.363 & 0.374 & 0.386 & 880 & 0.364 & 0.375 & 0.387 & 511 & 0.623 & 0.647 & 0.673\\
1708 & 252 & 852 & 0.283 & 0.294 & 0.306 & 903 & 0.268 & 0.278 & 0.288 & 450 & 0.530 & 0.558 & 0.587\\
1711 & 6 & 725 & 0.005 & 0.006 & 0.011 & 257 & 0.013 & 0.023 & 0.040 & 139 & 0.024 & 0.045 & 0.078\\
1714 & 128 & 698 & 0.169 & 0.182 & 0.196 & 631 & 0.185 & 0.201 & 0.219 & 160 & 0.746 & 0.796 & 0.841\\
\hline
\end{tabular}
\label{Tab:3y}
\end{table*}

\acknowledgement{
Support from the Junta de Extremadura (Research Group Grant No. GR10131), from the Ministerio de Econom\'ia y Competitividad
 of the Spanish Government (AYA2011-25945) and from the COST Action ES1005 TOSCA (http://www.tosca-cost.eu) is gratefully acknowledged.
I.U. and G.K. acknowledge support from ReSoLVE Centre of Excellence (Academy of Finland, project no. 272157).}


\begin{thebibliography}{31}
\expandafter\ifx\csname natexlab\endcsname\relax\def\natexlab#1{#1}\fi

\bibitem[{Beer {et~al.}(1998)Beer, Tobias, \& Weiss}]{beer98}
Beer, J., Tobias, S., \& Weiss, N. 1998, Solar Phys., 181, 237

\bibitem[{{Carrasco} {et~al.}(2015){Carrasco}, {Villalba {\'A}lvarez}, \&
  {Vaquero}}]{carrasco15}
{Carrasco}, V.~M.~S., {Villalba {\'A}lvarez}, J., \& {Vaquero}, J.~M. 2015,
  Solar Phys., submitted

\bibitem[{Casas {et~al.}(2006)Casas, Vaquero, \& Vazquez}]{casas06}
Casas, R., Vaquero, J., \& Vazquez, M. 2006, Solar Phys., 234, 379

\bibitem[{Clette {et~al.}(2014)Clette, Svalgaard, Vaquero, \&
  Cliver}]{clette14}
Clette, F., Svalgaard, L., Vaquero, J., \& Cliver, E. 2014, Space Sci. Rev.,
  this volume

\bibitem[{Eddy(1976)}]{eddy76}
Eddy, J. 1976, Science, 192, 1189

\bibitem[{Hoyt \& Schatten(1998a)}]{hoyt98}
Hoyt, D.~V. \& Schatten, K.~H. 1998a, Solar Phys., 179, 189

\bibitem[{{Hoyt} \& {Schatten}(1998b)}]{hoyt98a}
{Hoyt}, D.~V. \& {Schatten}, K.~H. 1998b, Solar Phys., 181, 491

\bibitem[{{Kovaltsov} {et~al.}(2004){Kovaltsov}, {Usoskin}, \&
  {Mursula}}]{kovaltsov04}
{Kovaltsov}, G.~A., {Usoskin}, I.~G., \& {Mursula}, K. 2004, Solar Phys., 224,
  95

\bibitem[{Manfredi(1736)}]{manfredi1736}
Manfredi, E. 1736, De Gnomone Meridiano Bononiensi ad Divi Petronii (Bononiae:
  Laeli a Vulpa), {397 pp}

\bibitem[{Mendoza(1997)}]{mendoza97}
Mendoza, B. 1997, Annales Geophys., 15, 397

\bibitem[{{Miyahara} {et~al.}(2010){Miyahara}, {Kitazawa}, {Nagaya},
  {Yokoyama}, {Matsuzaki}, {Masuda}, {Nakamura}, \& {Muraki}}]{miyahara10}
{Miyahara}, H., {Kitazawa}, K., {Nagaya}, K., {et~al.} 2010, J. Cosmol., 8,
  1970

\bibitem[{{Miyahara} {et~al.}(2004){Miyahara}, {Masuda}, {Muraki}, {Furuzawa},
  {Menjo}, \& {Nakamura}}]{miyahara04}
{Miyahara}, H., {Masuda}, K., {Muraki}, Y., {et~al.} 2004, Solar Phys., 224,
  317

\bibitem[{Miyahara {et~al.}(2006)Miyahara, Sokoloff, \& Usoskin}]{miyahara06}
Miyahara, H., Sokoloff, D., \& Usoskin, I. 2006, in Advances in Geosciences,
  Vol. 2: Solar Terrestrial (ST), ed. W.-H. Ip \& M.~Duldig (Singapore;
  Hackensack, U.S.A.: World Scientific), 1--20

\bibitem[{{Miyake} {et~al.}(2013){Miyake}, {Masuda}, \& {Nakamura}}]{miyake13}
{Miyake}, F., {Masuda}, K., \& {Nakamura}, T. 2013, Nature Comm., 4, 1748

\bibitem[{{Nagaya} {et~al.}(2012){Nagaya}, {Kitazawa}, {Miyake}, {Masuda},
  {Muraki}, {Nakamura}, {Miyahara}, \& {Matsuzaki}}]{nagaya12}
{Nagaya}, K., {Kitazawa}, K., {Miyake}, F., {et~al.} 2012, Solar Phys., 280,
  223

\bibitem[{{Owens} {et~al.}(2012){Owens}, {Usoskin}, \& {Lockwood}}]{owens12}
{Owens}, M.~J., {Usoskin}, I., \& {Lockwood}, M. 2012, Geophys. Res. Lett., 39,
  {L19102}

\bibitem[{Ribes \& Nesme-Ribes(1993)}]{ribes93}
Ribes, J. \& Nesme-Ribes, E. 1993, Astron. Astrophys., 276, 549

\bibitem[{{Roth} \& {Joos}(2013)}]{roth13}
{Roth}, R. \& {Joos}, F. 2013, Clim. Past, 9, 1879

\bibitem[{Soon \& Yaskell(2003)}]{soon03}
Soon, W.-H. \& Yaskell, S. 2003, The Maunder Minimum and the Variable Sun-Earth
  Connection (Singapore; River Edge, U.S.A.: World Scientific)

\bibitem[{Stuiver {et~al.}(1998)Stuiver, Reimer, Bard, Burr, Hughen, Kromer,
  McCormac, v.d. Plicht, \& Spurk}]{stuiver98}
Stuiver, M., Reimer, P., Bard, E., {et~al.} 1998, Radiocarbon, 40, 1041

\bibitem[{Usoskin {et~al.}(2001)Usoskin, Mursula, \&
  Kovaltsov}]{usoskin_JGR_MM_01}
Usoskin, I., Mursula, K., \& Kovaltsov, G. 2001, J. Geophys. Res., 106, 16039

\bibitem[{{Usoskin}(2013)}]{usoskin_LR_13}
{Usoskin}, I.~G. 2013, Liv. Rev. Solar Phys., 10, 1

\bibitem[{{Vaquero}(2007)}]{vaquero_rev_07}
{Vaquero}, J.~M. 2007, Adv. Space Res., 40, 929

\bibitem[{{Vaquero} {et~al.}(2011){Vaquero}, {Gallego}, {Usoskin}, \&
  {Kovaltsov}}]{vaquero11}
{Vaquero}, J.~M., {Gallego}, M.~C., {Usoskin}, I.~G., \& {Kovaltsov}, G.~A.
  2011, Astrophys. J. Lett., 731, {L24}

\bibitem[{{Vaquero} {et~al.}(2014){Vaquero}, {Guti{\'e}rrez-L{\'o}pez}, \&
  {Szelecka}}]{vaquero_ASR_14}
{Vaquero}, J.~M., {Guti{\'e}rrez-L{\'o}pez}, S., \& {Szelecka}, A. 2014, Adv.
  Space Res., 53, 1180

\bibitem[{Vaquero {et~al.}(2015)Vaquero, Nogales, \&
  Sánchez-Bajo}]{vaquero_ASR_15}
Vaquero, J.~M., Nogales, J.~M., \& Sánchez-Bajo, F. 2015, Adv. Space Res., 55,
  1546

\bibitem[{{Vaquero} \& {Trigo}(2014)}]{vaquero_SP_14}
{Vaquero}, J.~M. \& {Trigo}, R.~M. 2014, Solar Phys., 289, 803

\bibitem[{{Vaquero} {et~al.}(2012){Vaquero}, {Trigo}, \& {Gallego}}]{vaquero12}
{Vaquero}, J.~M., {Trigo}, R.~M., \& {Gallego}, M.~C. 2012, Solar Phys., 277,
  389

\bibitem[{{Vaquero} \& {V{\'a}zquez}(2009)}]{vaquero09}
{Vaquero}, J.~M. \& {V{\'a}zquez}, M. 2009, Astrophys. Space Sci. Lib., Vol.
  361, {The Sun Recorded Through History: Scientific Data Extracted from
  Historical Documents} (Berlin: Springer)

\bibitem[{Waldmeier(1961)}]{waldmeier61}
Waldmeier, M. 1961, The Sunspot Activity in the Years 1610-1960 (Z\"urich:
  Zurich Schulthess and Company AG)

\bibitem[{{Zolotova} \& {Ponyavin}(2015)}]{zolotova15}
{Zolotova}, N.~V. \& {Ponyavin}, D.~I. 2015, Astrophys. J., 800, 42

\end{thebibliography}

\end{document}